\documentclass[12pt]{article}

\usepackage{epsf,epsfig,subfigure}
\usepackage{amsmath,amsthm,amssymb,mathrsfs}
\usepackage{graphics}
\usepackage{graphicx}
\usepackage{indentfirst}
\usepackage[resetlabels]{multibib}

\makeatletter
\def\@biblabel#1{#1.}
\def\@cite#1#2{\textsuperscript{{#1\if@tempswa , #2\fi}}}
\makeatother

\def\centereps#1#2#3{\vskip#2\relax\centerline{\hbox to#1{\special
  {eps:#3 x=#1, y=#2}\hfil}}}

\def\centerbmp#1#2#3{\vskip#2\relax\centerline{\hbox to#1{\special
  {bmp:#3 x=#1, y=#2}\hfil}}}

\newcommand{\R}{\mathcal{R}}

\title{\bf Prevention and Control of Zika as a Mosquito-Borne and Sexually Transmitted Disease: A Mathematical Modeling Analysis}

\author
{{\bf Daozhou Gao}$^{1}$, {\bf Yijun Lou}$^{2}$, {\bf Daihai He}$^{2}$, {\bf Travis C. Porco}$^{3}$, \\
{\bf Yang Kuang}$^{4}$, {\bf Gerardo Chowell}$^{5}$, {\bf Shigui Ruan}$^{6, \ast}$\\
\\
\normalsize{$^{1}$Mathematics and Science College, Shanghai Normal University, Shanghai 200234, China}\\
\normalsize{$^2$Department of Applied Mathematics, The Hong Kong Polytechnic University,}\\
\normalsize{Hung Hom, Kowloon, Hong Kong, China}\\
\normalsize{$^3$Francis I Proctor Foundation, Department of Epidemiology and Biostatistics, }\\
\normalsize{and Department of Ophthalmology; University of California at San Francisco,}\\
\normalsize{San Francisco, CA 94143 USA}\\
\normalsize{$^4$School of Mathematical and Statistical Sciences, Arizona State University,}\\
\normalsize{Tempe, AZ 85287, USA}\\
\normalsize{$^5$School of Public Health, Georgia State University, Atlanta, GA 30302, USA}\\
\normalsize{$^6$Department of Mathematics, University of Miami, Coral Gables, FL 33146, USA}\\
\normalsize{$^\ast$}
Correspondence author. E-mail:  ruan@math.miami.edu.
}
\begin{document}
\baselineskip24pt
\maketitle

\newpage
%\begin{abstract}
\noindent
{\bf The ongoing Zika virus (ZIKV) epidemic poses a major global public health emergency. It is known that ZIKV is spread by \textit{Aedes} mosquitoes, recent studies show that ZIKV can also be transmitted via sexual contact and cases of sexually transmitted ZIKV have been confirmed in the U.S., France, and Italy. How sexual transmission affects the spread and control of ZIKV infection is not well-understood. We presented a mathematical model to investigate the impact of mosquito-borne and sexual transmission on spread and control of ZIKV and used the model to fit the ZIKV data in Brazil, Colombia, and El Salvador. Based on the estimated parameter values, we calculated the median and confidence interval of the basic reproduction number $\R_0=2.055$ (95\% CI: 0.523-6.300), in which the distribution of the percentage of contribution by sexual transmission is $3.044$ (95\% CI: 0.123-45.73). Our study indicates that $\R_0$ is most sensitive to the biting rate and mortality rate of mosquitoes while sexual transmission increases the risk of infection and epidemic size and prolongs the outbreak. In order to prevent and control the transmission of ZIKV, it must be treated as not only a mosquito-borne disease but also a sexually transmitted disease. }
%\end{abstract}

\newpage
%\section*{Introduction}
\noindent
Zika virus (ZIKV), a \textit{Flavivirus} closely related to dengue, is primarily transmitted to humans by the bites of infected female mosquitoes from the {\it Aedes} genus.  These mosquitoes, widespread in tropical and subtropical regions, also transmit dengue fever, chikungunya, yellow fever, and Japanese encephalitis. For ZIKV, about one in five infected people develops symptoms including mild fever, rash, conjunctivitis and joint pain, with no documented fatalities seen in a recent large outbreak \cite{duffy2009zika}.  It has been confirmed that ZIKV causes microcephaly in newborn babies of infected mothers \cite{Mlakar2016, Cauchemez2016} and some evidence suggests that it causes Guillain-Barr\'{e} syndrome (GBS) as well \cite{Cao-Lormeau2016}. Unfortunately, no vaccine, specific treatment, or fast diagnostic test is available to treat,  prevent, or diagnose ZIKV infection at this time.

The virus was initially isolated from a rhesus monkey in the Zika forest of Uganda in 1947, and later isolated from humans in Nigeria in 1954 \cite{dick1952zika,macnamara1954zika,Shapshak2015}.  Subsequently, only sporadic confirmed human cases were reported from Africa and Southeast Asia.  In April 2007, the first documented ZIKV outbreak outside traditionally affected areas occurred on Yap Island, Federated States of Micronesia, in the North Pacific \cite{duffy2009zika}.  In October 2013, a severe ZIKV outbreak was reported in French Polynesia, South Pacific, with an estimated 28,000 cases \cite{musso2014rapid}.  The ongoing outbreak, which began in April 2015 in Brazil, has rapidly spread to many other countries in South and Central America and the Caribbean with more than 140,000 suspected and confirmed cases by the end of February 2016 \cite{PAHO2016}. Nearly 6,000 suspected cases of microcephaly (including 139 deaths) among newborns might be linked to ZIKV infections in Brazil between October 2015 and February 2016. From December  2015  to  February  2016,  more than 200 GBS cases with history of suspected ZIKV infection were recorded in Colombia and
118  GBS (including 5 deaths) cases were reported in El  Salvador \cite{PAHO2016}.
The WHO declared the epidemic a Public Health Emergency of International Concern (PHEIC) on February 1, 2016 \cite{WHO2016}, and the U.S. CDC's Emergency Operations Center has moved to the highest level of activation on February 3, 2016 \cite{CDC2016}.
Based on the reported dengue data from 2015, WHO estimated that up to four million people in the Americas could be infected by ZIKV in 2016.  Without effective intervention, the situation has considerable potential to worsen, due in part to the upcoming 2016 Summer Olympics in Rio de Janeiro as well as anticipated mosquito abundance increases caused by an ongoing El Ni\~{n}o.

ZIKV has been detected in serum, saliva, urine, and semen \cite{gourinat2015detection, musso2015detection, Atkinson2016}.  It has also been detected in urine and semen even after it disappears from blood \cite{musso2015potential} and in one convalescent case it was detected in semen 27 and 62 days after onset of febrile illness \cite{Atkinson2016}.
Recent studies show that ZIKV can be transmitted via sexual contact.
%\cite{foy2011probable,musso2015potential},
%blood transfusion \cite{musso2014potential}, and perinatal infection \cite{besnard2014evidence}.
It was reported that an infected male had infected a female by having vaginal sexual intercourse, even before his onset of symptoms \cite{foy2011probable}.  After the confirmation of the first case of sexually transmitted ZIKV of the current outbreak in Dallas County by the CDC on February 2, 2016 \cite{dchhs2016first}, six more confirmed and probable cases of sexual transmission of ZIKV in the U.S. were reported by CDC on February 26, 2016 \cite{Hills2016}, and Europe's first case of sexually transmitted ZIKV was diagnosed in France in Febraury 2016 \cite{Mansuy2016}. 
A case of ZIKV infection imported in Florence, Italy ex-Thailand, leading to a secondary
autochthonous case, probably through sexual transmission in May 2014 was retrospectively diagnosed in 2016 \cite{Venturi2016}. 

%The presence of the virus in nasopharynx and urine underscores the possibility of person-to-person %transmission in a close setting \cite{fonseca2014first}.
%Musso et al. \cite{musso2014potential} detected ZIKV in 42 of 1,505 blood donors who were
%asymptomatic at the time of blood donation, of whom 11 had a Zika fever-like syndrome later on %\cite{musso2014potential}.

%These are consistent with the outcome of transmission by boar semen contaminated with Japanese %encephalitis virus (another {\it flavivirus}) to inseminated sows \cite{habu1977disorder} and the report of %dengue virus transmission through mucosal surfaces \cite{chen2004transmission}.

The study of the ZIKV outbreak on Yap Island \cite{duffy2009zika} indicates that cases occured among all age groups, but the incidence of ZIKV disease was highest among persons 55 to 59 years of age with the mean age of 36 years and 61\% female. Since ZIKV infections are mostly asymptomatic or have mild symptoms lasting two to seven days, the disease has little impact on sexual activity \cite{Hills2016}.  If ZIKV is sexually transmissible, then it is necessary to abstain from sexual activity or consistently use condoms during convalescence and the CDC issued interim guidance on safe sex during a Zika outbreak \cite{CDC2016b}.  This is particularly important to pregnant women in areas where the ZIKV is circulating.

\section*{Results}
\begin{figure}[htbp]
\centering
\includegraphics[width=5.0in]{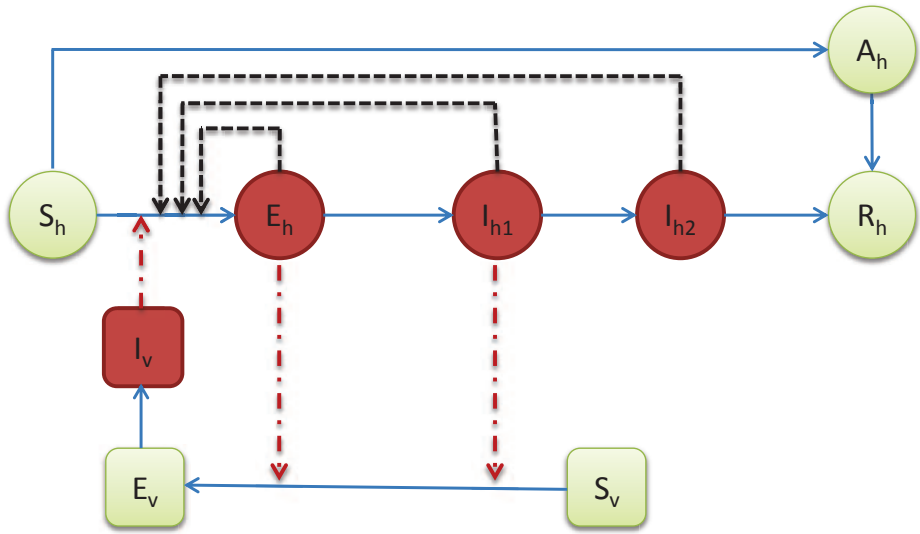}
\caption{Flow diagram for the spread of ZIKV involving vectorial and sexual transmission.  Green nodes are non-infectious and red nodes are infectious.  Blue solid arrows show the progression of infection.  Black dashed arrows show direction of human-to-human transmission and red dash-dotted lines show direction of transmission between humans and mosquitoes.  An individual may progress from susceptible ($S_h$) to asymptomatically infected ($A_h$) to recovered ($R_h$), or exposed ($E_h$) to symptomatically infected ($I_{h\!1}$) to convalescent ($I_{h\!2}$) to recovered ($R_h$), while a mosquito may progress from susceptible ($S_v$) to exposed ($E_v$) to infectious ($I_v$). }
\label{flowchart}
\end{figure}

\noindent
{\bf Modeling}. \
Mathematical modeling has become a crucial tool in desigining prevention and control measures for infectious diseases \cite{anderson1991, KeelingRohani}.
To investigate the role of sexual transmission in the spread and control of Zika virus disease, we developed a deterministic model of Zika disease transmission that takes into account both mosquito-borne and sexual transmission modes  (Fig. \ref{flowchart}).
%Human population is divided into five classes: susceptible ($S_h(t)$), exposed ($E_h(t)$), symptomatic %infectious ($I_{h\!1}(t)$), convalescent ($I_{h\!2}(t)$), and recovered ($R_h(t)$) at time $t>0,$ and the %mosquito population is divided into three classes: susceptible ($S_v(t)$), exposed ($E_v(t)$), and %infectious ($R_v(t)$), respectively.
%$N_h=S_h+E_h+I_{h\!1}+I_{h\!2}+R_h$ denote the total number of humans and $N_v=S_v+E_v+I_v$ %denote the total mosquito population, which are assumed to be constant.
Symptomatically infected humans are contagious to both mosquitoes and humans during the incubation period that is typically between 2 and 7 days. This is because the viremia and virusemenia occur before the end of the incubation period, although the viral load of exposed (presymptomatic) people may be lower \cite{foy2011probable}.  After this period, infected humans develop symptoms. Symptomatic humans are more contagious to mosquitoes than
exposed individuals and are also able to transmit the virus to partners through sex \cite{Hills2016}.  The virus appears to persist longer in semen and urine than in serum \cite{musso2015potential, Atkinson2016}. Following the period of viremia, symptomatic humans enter the convalescent stage and can no longer infect mosquitoes.  However, such individuals remain infectious to humans, though with reduced infectivity.  The infected humans' convalescent period ends with
lifelong immunity. Sexual transmission of ZKIV from asymptomatically infected humans has not been documented, so they are assumed to be noninfectious to humans.  The timescale of human demography is far longer than that of the epidemiological dynamics, so we ignore human births and deaths when modeling an outbreak.

We make the following additional assumptions: (i) Mosquitoes cannot be infected by biting asymptomatically ZIKV infected people; (ii) The sexual ratio of humans is 1:1 and male and female are subject to almost the same epidemiological factors; (iii) The end of the viremic period coincides with the disappearance of symptoms in symptomatically infected individuals (see Fig. \ref{flowchart}).

\newpage
%\vspace*{0.3cm}
\noindent
{\bf Parameter estimates}. \
All parameter descriptions and ranges are summarized in Table \ref{table1}. 
To parameterize our model, we used reasonable epidemiological parameters based on our current understanding of Zika epidemiology and transmission dynamics.
Zika virus and dengue virus are arboviruses of the same genus {\it Flavivirus} spread by mosquitoes of the same genus {\it Aedes} and have similar symptoms, high proportion of asymptomatic infections, duration of incubation and infectiousness \cite{Shapshak2015}. Hence they have the same number of bites on humans per mosquito per unit time, $a$, and we anticipated that their transmission probabilities per bite from mosquitoes to humans, $b$, and from humans to mosquitoes, $c$, respectively, are comparable. 

Many parameters of ZIKV infection are not available, but we can make some reasonable assumptions to estimate specific measurements of the infectivity.  For instance, consideration of other human sexually transmitted infections may provide an initial basis for analysis.  We note that gonorrhea appears to display a high transmission probablity of transmission per coital act (results ranging from 0.19--0.65, with male to female transmission higher than female to male \cite{craig-gray-edwards2015}).
Others show a considerably lower transmission probability per coital act, such as
% {\it Chlamydia trachomatis} (0.095) \cite{althaus2012towards}.
HSV-2 (0.0005) \cite{wald2001effect}.
%Here we assume that the sexual transmission probablity rages from $0.001$ to $0.10.$
%and HPV (0.004) \cite{burchell2011genital}.
%recommend omit HIV
%Studies of HIV revealed a low estimated probability of transmission per contact
%(approximately 0.001) \cite{gray2001probability,hughes2012determinants,patel2014estimating}.
%Substantial heterogeneity was seen, with per-partnership transmission probabilities of 0.1
%reported early in the epidemic \cite{Grant1987}.

The average sexual frequency over sexually active ages is twice a week and the frequency of sexual intercourse over all age groups is assumed to be once a week \cite{laumann1994social}.  The sexual transmission rate of symptomatically infected people (transmission probability $\times$ contact rate), $\beta$, is assumed to range from 0.001 to 0.10, which means the transmission probability per sex act is between 0.007 (mild infectivity) and 0.70 (severe infectivity).

\begin{table}[htbp]
\caption{Parameter descriptions and ranges of the model (time unit is day). }
\vspace*{0.5cm}
\begin{center}
\vspace*{-0.6cm}
\begin{tabular}{llllll}
%\toprule
\hline\noalign{\smallskip}
& Description & Range    & Value & Reference \\
%\midrule
\noalign{\smallskip}\hline\noalign{\smallskip}
$a$: & mosquito biting rate &   0.3--1    & 0.5  & \cite{andraud2012dynamic} \\

$b$: & transmission probability from infectious&    0.1--0.75   & 0.4 & \cite{andraud2012dynamic}  \\
  &  mosquitoes to susceptible humans  &       &  &  \\

$c $: & transmission probability from symptomatically&  0.3--0.75     & 0.5 &  \cite{chikaki2009dengue}\\
  &  infected humans to susceptible mosquitoes &       &  &  \\

$\eta$: & relative human-to-mosquito transmission &    0--0.3   & 0.1 & Assumed  \\
  &   probability of exposed humans to    &       &  &  \\
  &   symptomatically infected humans  &       &  &  \\

$\beta$: & transmission rate from symptomatically &   0.001--0.10   & 0.05 & Assumed \\
  &  infected humans to susceptible humans  &       &  &  \\

$\kappa$: & relative human-to-human transmissibility of   &   0--1   &  0.6 & Assumed \\
  &  exposed humans to symptomatic humans &       &  &  \\

$\tau$: & relative human-to-human transmissibility of   &   0--1   &  0.3 & Assumed \\
  &  convalescent humans to symptomatic humans &       &  &  \\

$\theta$(\%): & proportion of symptomatic infection & 10--27    &  18 &  \cite{duffy2009zika} \\

$m$: & average ratio of mosquitoes to humans & 1--10    & 5  &  \cite{de2011modeling} \\

$1/\nu_h$: & intrinsic incubation period in humans & 2--7  &    5   &  \cite{bearcroft1956zika} \\  %3-12

$1/\nu_v$: & extrinsic incubation period in mosquitoes & 8--12   & 10  & \cite{andraud2012dynamic,boorman1956simple} \\

$1/\gamma_{h\!1}$: & duration of acute phase & 3--7     &  5  & \cite{bearcroft1956zika} \\ %2-7

$1/\gamma_{h\!2}$: & duration of convalescent phase & 14--30    & 20  & \cite{gourinat2015detection,musso2015potential} \\

$1/\gamma_h$: & duration of asymptomatic infection & 5--10 &  7  & Assumed \\

$1/\mu_v$: & mosquito lifespan & 4--35 &  14  & \cite{andraud2012dynamic,chikaki2009dengue} \\
%\bottomrule
\noalign{\smallskip}\hline
\end{tabular}
\label{table1}
\end{center}
\end{table}

\begin{figure}[h!]
(A) \centerline{\includegraphics[width=15cm, height=9cm]{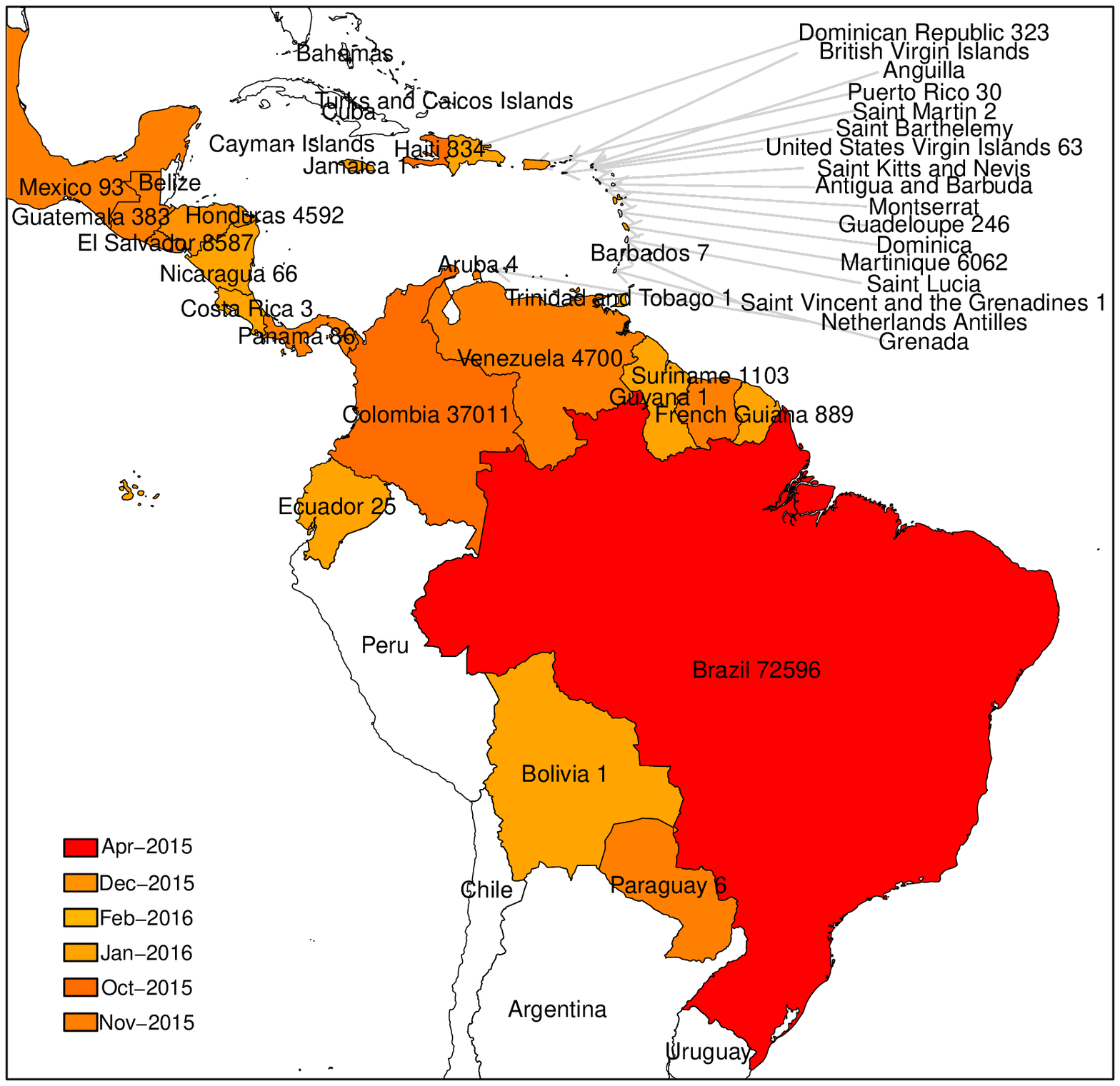}}\\
(B) \includegraphics[width=16cm]{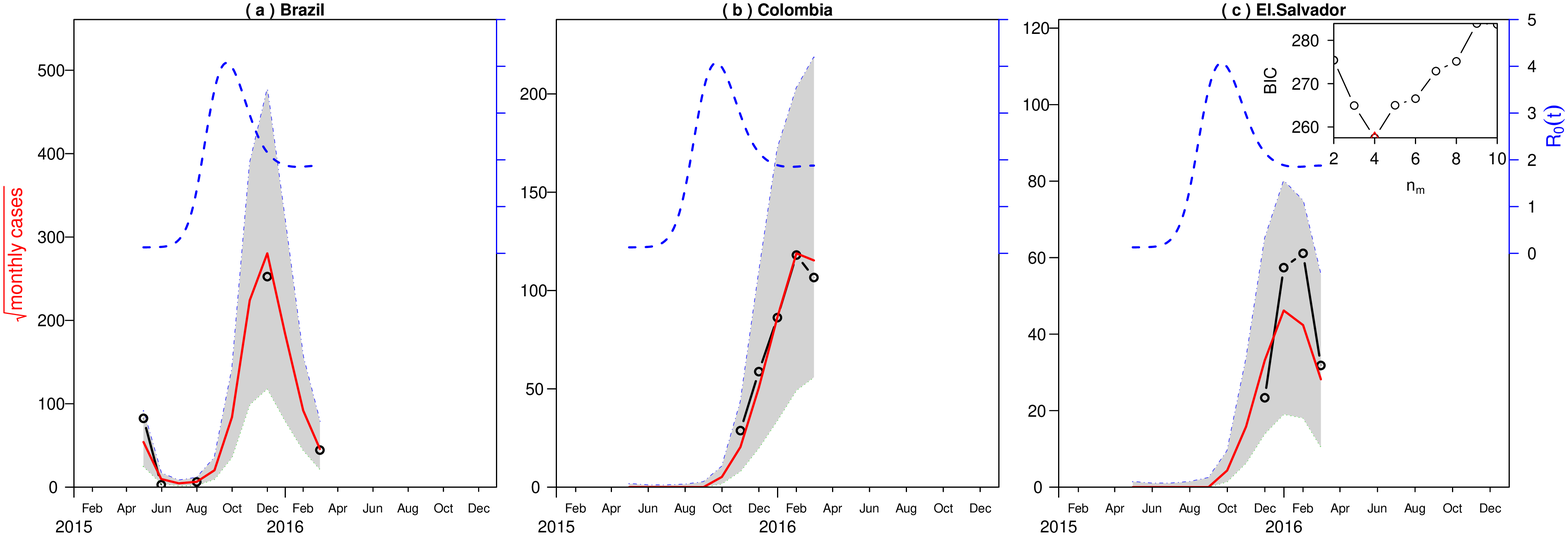}
\caption{(A) ZIKV outbreaks in South and Central Americas. The map indicates the month of first reported cases and the cumulative cases by February 27, 2016, in each country. The map was made with the free software R (http://www.r-project.org) and the country borders were from Sandvik, B. World Borders Dataset, http://thematicmapping.org (2009), accessed on February 1, 2016. (B) Fitting model to data in Brazil, Colombia, and El Salvador. Each panel shows the simulation (red solid curve) versus the observed (black circle), with the best fitting parameters. The red solid curves show median values of 1000 simulations and shaded region show the 95\% range. The blue dash curves show the estimated mosquito-human population ratio $m(t)$.  The inset panel shows Bayesian Information Criterion (BIC) as a function of the number of nodes ($n_m$) in $m(t)$ with values $m_i$ at these nodes. Assumed or estimated parameters and initial conditions are given in Table \ref{table2}.}
\label{fitting9}
\end{figure}

\vspace*{0.3cm}
\noindent
{\bf Fitting Zika data in Brazil, Colombia and El Salvador}. \ 
To use our model to fit the reported ZIKV cases up to February 27, 2016, in Brazil, Colombia, and El Salvador (see Fig. \ref{fitting9}(A)), we assumed that the three countries share common parameter values (see Table \ref{table1}), except for country population size and initial conditions (see Table \ref{table2}). Since large scale mosquito-control campaign has been taken in these Zika affected countries, we assumed that the ratio of mosquitoes to humans $m$ is time-dependent and used a cubic spline function of time with $n_m$ parameters to describe $m(t).$ 

Fig. \ref{fitting9}(B) demonstrates that our model provides good fits to the reported Zika data from Brazil, Colombia, and El Salvador. Since $m(t)$ is time-depedent, so is $R_0(t)$ which is represented by the right vertical axis. In Brazil, the outbreak started in the spring of 2015, has passed its peak, and seems under control for the time being. In  Colombia and El Salvador, the disease started in the summer of 2015 and is reaching its peak now. More Zika, GBS and microcephaly cases are expected from other countries in South and Central Americans and the Caribbean. The starting time and geographic spread of Zika (Fig. \ref{fitting9}(A)) indicates that it is following the path of dengue and chikungunya
and has the potential to be introduced to many other countries where the {\it Aedes} species mosquitoes are competent,  including some southern states in the U.S.

\newpage
\begin{table}[htbp]
\caption{Parameter values and initial conditions used in Fig. \ref{fitting9}.}
\vspace*{0.5cm}
\begin{center}
\vspace*{-0.6cm}
%\begin{tabular}{llllll}
\begin{tabular}{l|ll} 
\hline
%\noalign{\smallskip}
Parameters & estimated & assumed range \\
\hline
$\rho$ & 0.012  & [0,1] \\
$\tau_1$ & 0.404  & [0,$\infty$]\\
$m_1$ &   0.001  & [0,20]\\
$m_2$ &   9.525  & [0,20]\\
$m_3$ &  13.100  & [0,20]\\
$m_4$ &   8.129  & [0,20]\\
$S_{h,\text{Brazil}}$ &    0.516  & [0.5,0.95]\\
$E_{h,\text{Brazil}}$ & 0.000657  & [9.79e-9,1]\\
$I_{h1,\text{Brazil}}$ & 0.000657  & $\equiv E_{h,\text{Brazil}}$ \\
$I_{h2,\text{Brazil}}$ & 0.000657  & $\equiv E_{h,\text{Brazil}}$ \\
$A_{h,\text{Brazil}}$ & 0.000657  & $\equiv E_{h,\text{Brazil}}$ \\
$S_{v,\text{Brazil}}$ &  & 1-2e-4\\
$E_{v,\text{Brazil}}$ &  & 1e-4\\
$I_{v,\text{Brazil}}$ &  & 1e-4 \\
$S_{h,\text{Colombia}}$ &    0.705  & [0.5,0.95]\\
$E_{h,\text{Colombia}}$ & 2.515e-07  & [4.27e-8, 2.13e-7]\\
$I_{h1,\text{Colombia}}$ & 2.515e-07  & $\equiv E_{h,\text{Colombia}}$ \\
$I_{h2,\text{Colombia}}$ & 2.515e-07  & $\equiv E_{h,\text{Colombia}}$ \\
$A_{h,\text{Colombia}}$ & 2.515e-07  & $\equiv E_{h,\text{Colombia}}$ \\
$S_{v,\text{Colombia}}$ &   & 1-2e8\\
$E_{v,\text{Colombia}}$ &   & 1e-8 \\
$I_{v,\text{Colombia}}$ &   & 1e-8 \\
$S_{h,\text{El Salvador}}$ &    0.647  & [0.5, 0.95]\\
$E_{h,\text{El Salvador}}$ & 8.076e-07  & [3.25e-7, 1.628e-6]\\
$I_{h1,\text{El Salvador}}$ & 8.076e-07  & $\equiv E_{h,\text{El Salvador}}$ \\
$I_{h2,\text{El Salvador}}$ & 8.076e-07  & $\equiv E_{h,\text{El Salvador}}$ \\
$A_{h,\text{El Salvador}}$ & 8.076e-07  & $\equiv E_{h,\text{El Salvador}}$ \\
$S_{v,\text{El Salvador}}$ &   & 1-2e-8\\
$E_{v,\text{El Salvador}}$ &   & 1e-8\\
$I_{v,\text{El Salvador}}$ &   & 1e-8 \\
\hline
%\noalign{\smallskip}\hline
\end{tabular}
\label{table2}
\end{center}
%\end{tabular}
\end{table}
\clearpage

\clearpage

\begin{figure}[htbp]
\centering
(A) \includegraphics[width=4in]{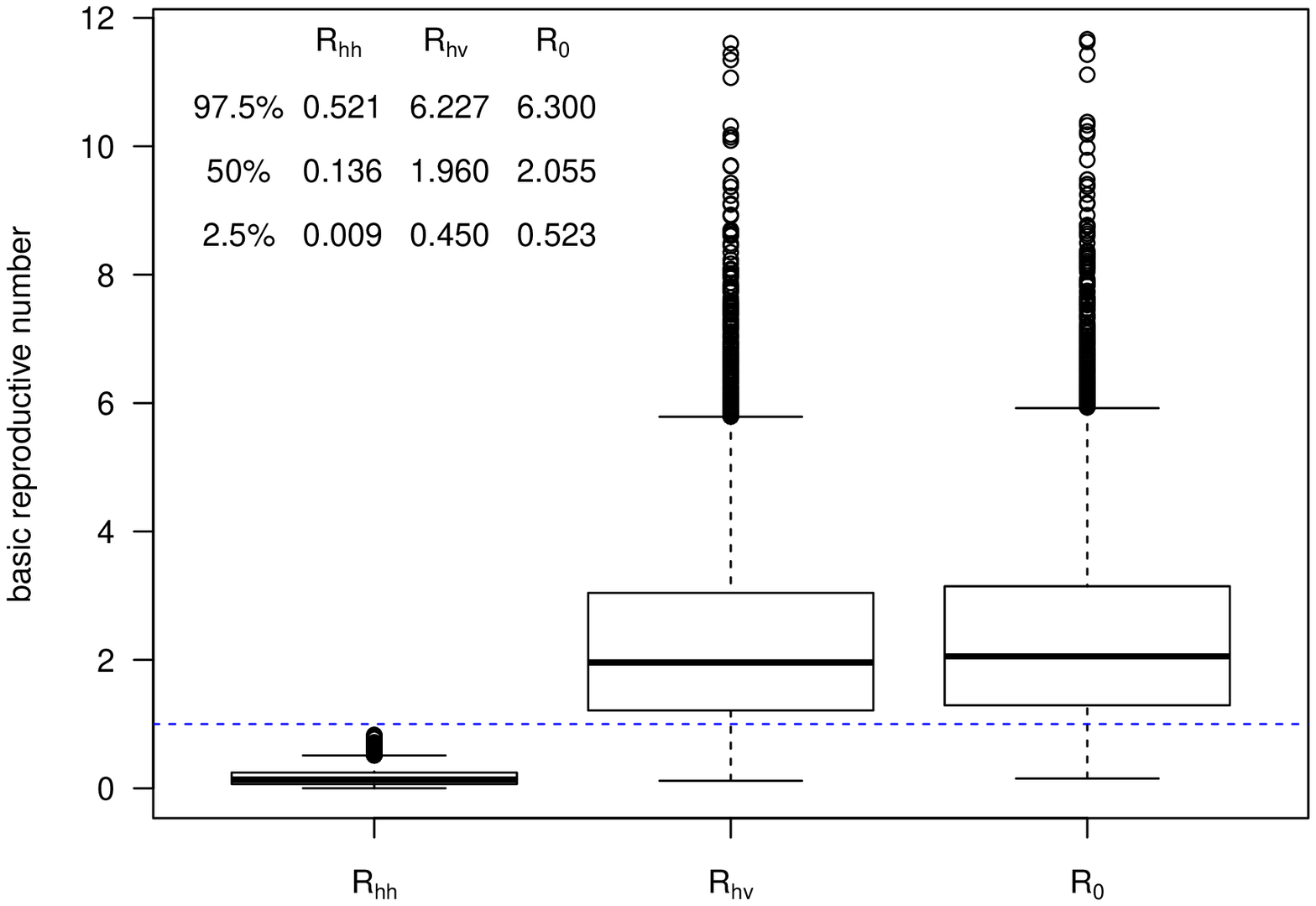}\\
(B) \includegraphics[width=4in]{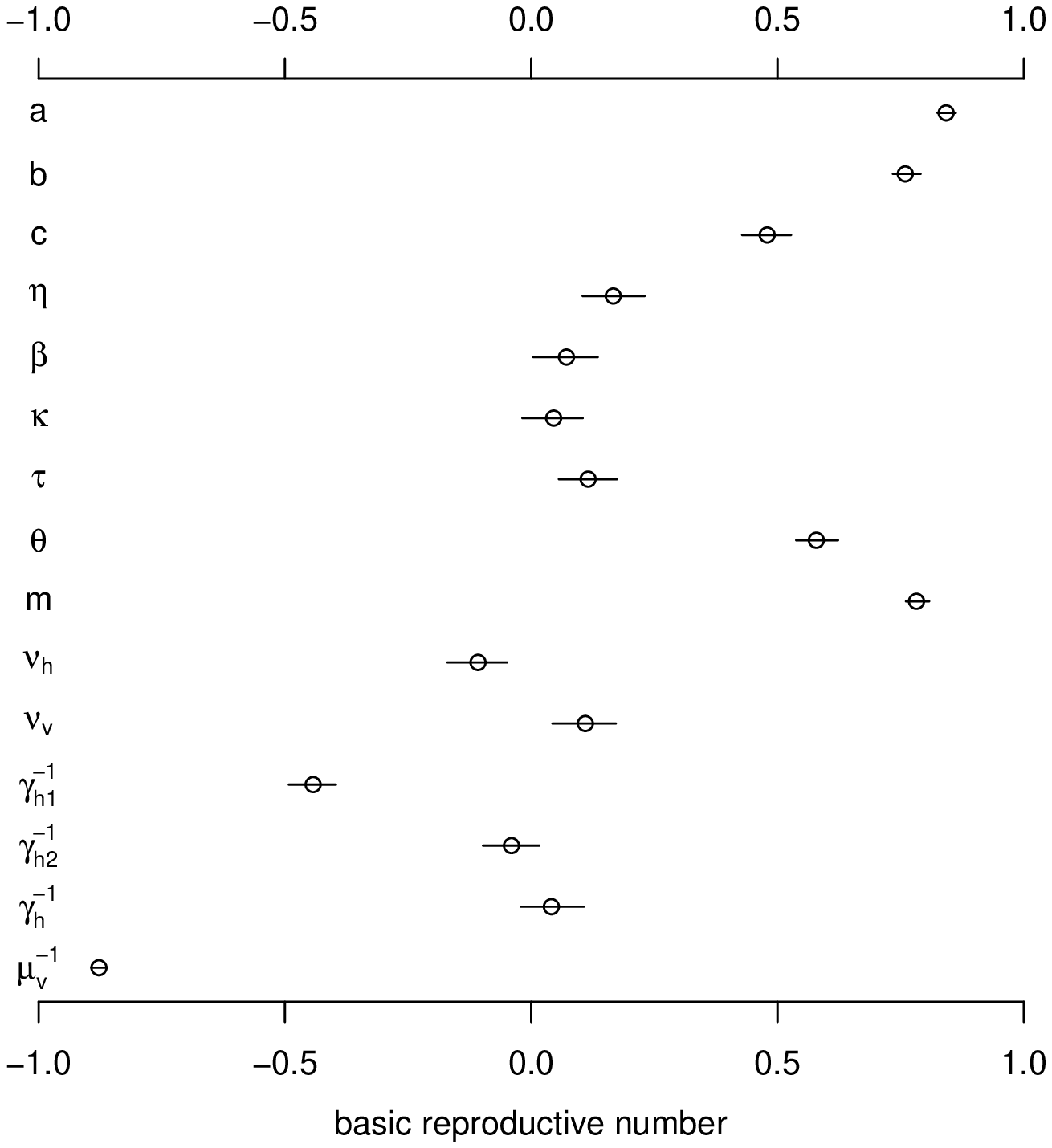}
\caption{(A) The box plot of the basic reproduction numbers for one or both transmission routes. The blue horizontal dashed line is $\R_0=1$. (B) The partial rank correlation coefficient (PRCC) of the basic reproduction number with respect to model parameters. The circle is the estimated correlation and the bar shows the 95\% confidence interval. Parameter ranges are given in Table \ref{table1}. }
\label{basicplot}
\end{figure}

\vspace*{0.3cm}
\noindent
{\bf Estimation of the basic reproduction number}. \ 
Based on parameter ranges in Table \ref{table1}, we used the Latin hypercube sampling method \cite{iman-helton-campbell1981} to generate 5,000 samples by assuming a uniform distribution for each parameter, and calculated the corresponding uncertainty on the basic reproduction numbers of either mosquito-borne transmission or sexual transmission or both.  The median and confidence interval of the distribution of the basic reproduction numbers (see Fig. \ref{basicplot}(A)) are 2.055 (95\% CI: 0.523-6.300) for $\R_0$, 1.960 (95\% CI: 0.450-6.227) for $\R_{hv}$, and 0.136 (95\% CI: 0.009-0.521) for $\R_{hh}$, respectively; the median and confidence interval of the distribution of the percentage of contribution by sexual transmission in $\R_0$ is 3.044 (95\% CI: 0.123-45.73).  This suggests that sexual transmission alone is unlikely to initiate or sustain an outbreak.  However, if the human-to-human transmission probability is very high, then its promoting effect on the transmission of ZIKV cannot be neglected.

To identify the key factors that determine the magnitude of the basic reproduction number, we performed global sensitivity analysis (see Fig. \ref{basicplot}(B)) with 1,000 random sample uniformly distributed in the range of parameter values from Table \ref{table1} and 1,000 bootstrap replicates \cite{saltelli2000sensitivity}.  The basic reproduction number is most sensitive to mosquito biting rate and mortality rate.

\clearpage

\begin{figure}[htbp]
\centering
\includegraphics[width=6.5in]{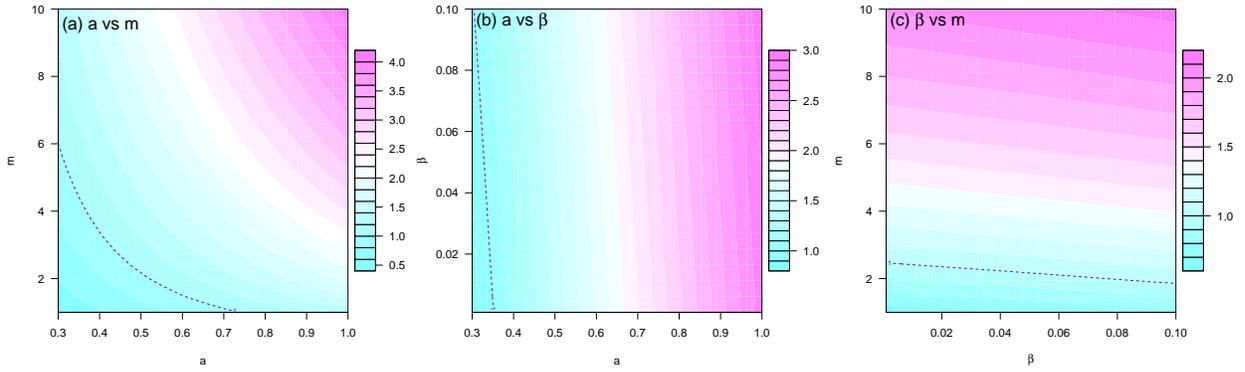}
%(B) \includegraphics[width=6in]{AR_ambeta.eps}
\caption{The contour plot of the basic reproduction number in terms of two of the three controllable parameters: $\beta$ (transmission rate of symptomatically infected humans to susceptible humans), $m$ (ratio of mosquitoes to humans), and $a$ (mosquito biting rate). The blue dashed curve is the contour of $\R_0=1$. Other parameter values are given in Table \ref{table1}.}
%(B) The contour plot of the attack rate in terms of two of the three controllable parameters: $\beta$, $m$, %and $a$. Other parameter values are given in Table \ref{table1}.}
\label{contour}
\end{figure}

To determine the dependence of the basic reproduction number on the controllable model parameters, choose $b=0.4, c=0.5, \eta=0.1, \kappa=0.6, \tau=0.3, \theta=0.18, \nu_h=1/5, \nu_v=1/10, \gamma_{h\!1}=1/5, \gamma_{h\!2}=1/20, \gamma_h=1/7, \mu_v=1/14$, and the three controllable parameters $\beta=0.05$ or $(0.001,0.10)$, $m=5$ or $(1,10)$, $a=0.5$ or $(0.3,1)$. Fixing one of $\beta, m$, and $a$ at the specific value and varying the other two parameters, the contour plots of $\R_0$ in terms of $a$ and $m$ (left panel), $a$ and $\beta$ (middle panel), and $\beta$ and $m$ (right panel) are illustrated in Fig. \ref{contour}, respectively. The basic reproduction number ranges from 0.5 to 4, the disease cannot spread if $\R_0<1$ and can cause an outbreak otherwise.

Fig. \ref{contour} indicates that mosquito-control, personal biting protection, and sexual contact protection are all important measures to control Zika virus infection. As season changes, $m$ will change along time which may cause fluctuating outbreaks of Zika in the future. Note that even the vector-control and bitting pretection measures are successful, if the human-to-human sexual transmission probablity is very high, then the disease still could persist in the population and the outbreak could be prolonged.

\newpage
\clearpage

\vspace*{0.3cm}
\noindent
{\bf Attack Rate}. \ 
Attack rate or attack ratio $z$ is the fraction of the population that becomes infected, which is also an important concept in measuring the transmission of infectious diseases.

\begin{figure}[htbp]
\begin{center}
\includegraphics[width=5in]{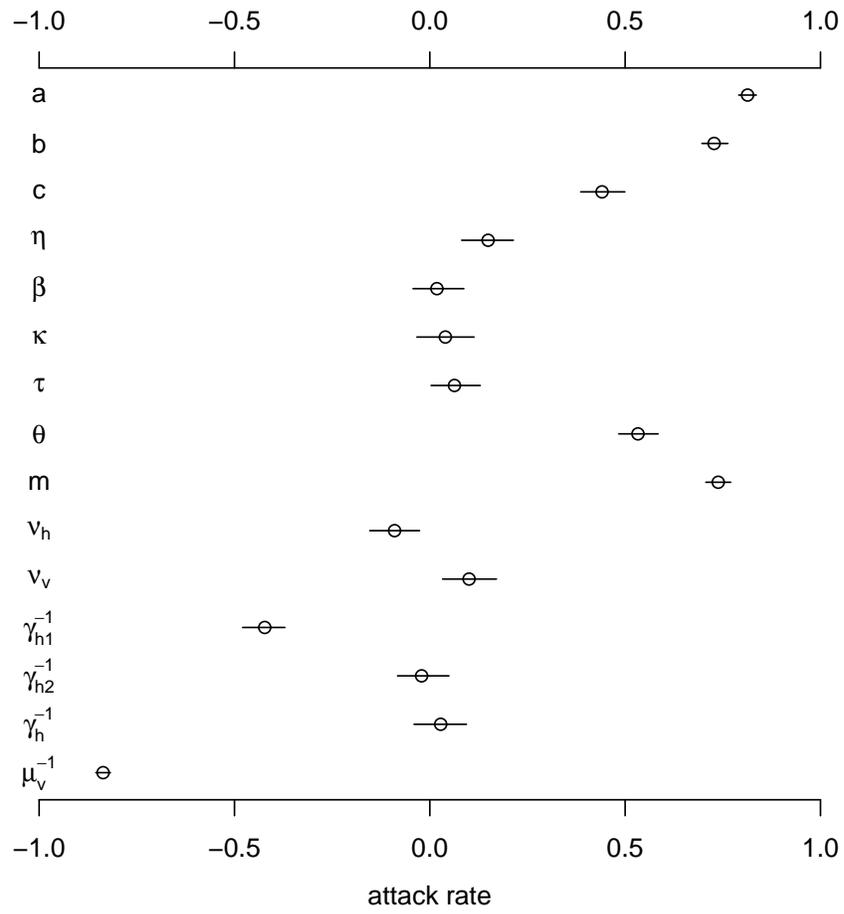}
\end{center}
\caption{The partial rank correlation coefficient (PRCC) of the attack rate with respect to model parameters. The circle is the estimated correlation and the bar shows the 95\% confidence interval. Parameter ranges are given in Table \ref{table1}.}
\label{figure5}
\end{figure}

To identify the key factors that affect the attack rate, we performed global sensitivity analysis (see Fig. \ref{figure5}) with 1,000 random sample uniformly distributed in the range of parameter values from Table \ref{table1} and 1,000 bootstrap replicates.  The attack rate is also most sensitive to mosquito biting rate and mortality rate.

\newpage
\clearpage

\begin{figure}[htbp]
\begin{center}
\includegraphics[width=6in]{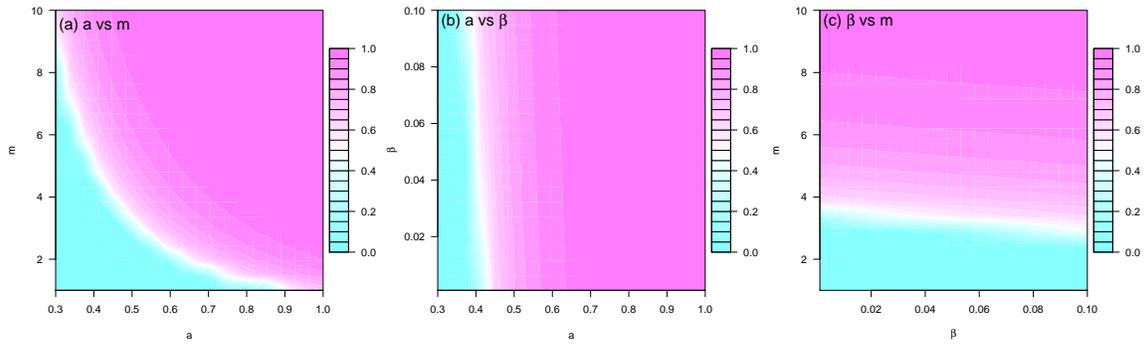}
\end{center}
\caption{The contour plot of the attack rate in terms of two of the three controllable parameters: $\beta$ (transmission rate of symptomatically infected humans to susceptible humans), $m$ (ratio of mosquitoes to humans), and $a$ (mosquito biting rate). All parameter values are given in Table \ref{table1}.}
 \label{figure6}
\end{figure}

To determine the dependence of the attack rate on the controllable model parameters, choose $b=0.4, c=0.5, \eta=0.1, \kappa=0.6, \tau=0.3, \theta=0.18, \nu_h=1/5, \nu_v=1/10, \gamma_{h\!1}=1/5, \gamma_{h\!2}=1/20, \gamma_h=1/7, \mu_v=1/14$, and the three controllable parameters $\beta=0.05$ or $(0.001,0.10)$, $m=5$ or $(1,10)$, $a=0.5$ or $(0.3,1)$. Fixing one of $\beta, m$, and $a$ at the specific value and varying the other two parameters, the contour plots of the attack rate $z$ in terms of $a$ and $m$ (left panel), $a$ and $\beta$ (middle panel), and $\beta$ and $m$ (right panel), are illustrated in Fig. \ref{figure6}, respectively.

Based on the attack rate 73\% (95\% CI: 68\%--77\%) observed in the Yap Island ZIKV outbreak of 2007 \cite{duffy2009zika}, we estimated the basic reproduction number according to the final size formula $R_0=-\log(1-f_{\infty})/f_{\infty}$ in the absence of behavior changes, control interventions, and so on, where $f_{\infty}$ is the final infected fraction \cite{ma2006generality}. These yield $R_0=1.79$ with uncertainty interval 1.68--1.91.  We further note that with 11\% of the population seeking medical care for suspected ZIKV infection during the French Polynesia ZIKV outbreak of 2013
\cite{musso2014rapid}, consistent with a total attack fraction over 0.55 and $R_0>1.4$ (serological surveys of the attack fraction in French Polynesia are ongoing). %AUBRY PERS COMM
\textbf{These estimates are roughly consonant with the value of $\R_0$ (1.6-2.5) for dengue in Brazil
\cite{marques1994basic}.}
Of course, the estimate of $R_0$ was obtained by assuming random mixing and a constant vector population over the course of the epidemic. Extrapolation of the attack rate seen in the Yap Island outbreak would be potentially misleading, due to differences in vectorial capacity, heterogeneity in space, as well as vector control and infection minimization strategies (e.g., bednets) that would occur during larger scale epidemics.

\newpage
\clearpage

\vspace*{0.3cm}
\noindent
{\bf Numerical Scenarios}. \
In Table \ref{table3}, selected scenarios from Fig. \ref{contour} are presenteded, demonstrating the
range of transmission contributed by sexual activity.  If $m=5$ and $a=0.5$, sexual transmission only accounts for 1.351\% of
total infections at $\beta=0.01$, increases to 6.506\% if $\beta=0.05$ and 12.45\% if $\beta=0.10$.
When $\beta=0.05, m=5, a=0.5$, the attack rate is up to 82.65\%; it is impossible to contain infection even reducing mosquito population by 55\%, though the reproduction number for human-mosquito transmission is less than one in
this scenario. In this case, both public and individual disease intervention measures are imperative (e.g., 55\% reduction in mosquito population and 50\% reduction in unprotected sex behaviors).
In all scenarios, $b=0.4, c=0.5, \eta=0.1, \kappa=0.6, \tau=0.3, \theta=0.18, \nu_h=1/5, \nu_v=1/10, \gamma_{h\!1}=1/5, \gamma_{h\!2}=1/20,\gamma_h=1/7, \mu_v=1/14$. The three columns refer to controllable parameters, reproduction numbers, and attack rates, respectively.

\newpage
\begin{table}[htbp]
\caption{Numerical scenarios for Fig. \ref{contour} and \ref{figure6}. $\R_{p}=\R_{hh}/(\R_{hh}+\R_{hv}^2)$ and $z_p=z_{hh}/z$ are the percentage of contribution by sexual transmission in the basic reproduction number and attack rate, respectively.}
\begin{center}
%\vspace*{-0.6cm}
\begin{tabular}{ccc|cccc|cccc}
\hline
$\beta$ & $m$ & $a$ & $\R_0$ & $\R_{hv}$ & $\R_{hh}$ & $\R_{p}$(\%)  & $z$(\%) & $z_{hv}$(\%) & $z_{hh}$(\%) & $z_{p}$(\%) \\
\hline

0.05 & 5 & 0.5 & 1.486 & 1.422 & 0.126 & 5.868  & 82.65 & 77.27 & 5.377 & 6.506 \\

0 & 5 & 0.5 & 1.422 & 1.422 & 0 &  0  & 79.60 & 79.60 & 0  & 0 \\

0.10 & 5 & 0.5 & 1.553 & 1.422 & 0.252  & 11.09  & 85.18 & 74.58 & 10.60 & 12.45 \\

0.05 & 2.25 & 0.5 & 1.019 & 0.954 & 0.126  & 12.17  & 6.782 & 5.957 & 0.825 & 12.17\\

0 & 2.25 & 0.5 & 0.954 & 0.954 & 0  &  0  & 0 & 0 & 0 & 0 \\

0.01 & 5 & 0.5 & 1.434 & 1.422 & 0.025  &  1.231  & 80.26 & 79.17 & 1.084  & 1.351 \\

0.01 & 10 & 0.5 & 2.023 & 2.011 & 0.025  & 0.620   & 97.89 & 97.02 & 0.872  & 0.890\\

0.01 & 5 & 1 & 2.856 & 2.843 & 0.025  &  0.311  & 99.94 & 99.28 & 0.663  & 0.663 \\

0.05 & 10 & 0.5 & 2.075 & 2.011 & 0.126   & 3.023  & 98.10 & 93.81 & 4.292 & 4.375\\

0.05 & 5 & 1 & 2.907 & 2.843 & 0.126   & 1.535  & 99.94 & 96.66 & 3.280 & 3.282\\

\hline
\end{tabular}
\label{table3}
\end{center}
\end{table}
\clearpage

\section*{Discussion}
\noindent
Lacking data on early dynamics to dissect the contributions from different transmission routes of ZIKV,  we designed a SEIR-type model based on classic epidemic theory \cite{anderson1991} and studied the impact of sexual transmission on the prevention and control of the mosquito-borne ZIKV. As a result, we  considered here the exponential growth dynamics of ZIKV infection. In contrast, sub-exponential growth dynamics for sexually-transmission diseases has been reported previously  \cite{Colgate1989}. \textbf{By using reasonable ranges of parameter values, we obtained some estimates of the basic reproduction number and provided a qualitative analysis of the contribution of mosquito-borne transmission and sexual transmission on the spread of ZIKV}.

%A simple mosquito-borne disease model with human-to-human transmission permitted an assessment of %the role of sexual transmission on the spread and control of Zika virus.
Sexual transmission increases the risk of infection and epidemic size, but itself may not initiate or sustain an outbreak.  Statistically, the transmission contributed by sexual activity is a small percentage of the total transmission in attack rate (4.437 (95\% CI: 0.297-23.02)).
%unless the human-to-human transmissibility is implausibly high.
However, the potential of non-vector-borne transmission could complicate efforts on containing the spread of ZIKV. Under certain circumstances, culling a large fraction of mosquito population
such that the basic reproduction number for human-mosquito transmission below one may be not sufficient in the absence of sexual risk reduction.

At the beginning of an outbreak, it is best to contain ZIKV infection by vector control (larviciding and adulticiding) and insect bite precautions (bed net and insect repellents).  However, insecticide resistance in {\it Aedes aegypti} and {\it Aedes albopictus}
and poverty in the most vulnerable regions can compromise the success of bite reduction.
%\cite{rawlins1995resistance,lima2003resistance,rodriguez2007levels,vontas2012insecticide}.
Subsequently, measures on reduction of sexual transmission shall be implemented. Especially, pregnant women or those at childbearing age who have pregnancy planning need to avoid unprotected sexual contact.

\textbf{To better understand the transmission mechanisms and estimate the level of sexual transmissibility, solid clinical and epidemiological data are required.} However, the cross-reaction of other {\it Flaviviruses} and the mild illness of Zika make the infections highly unreported or misdiagnosed. The magnitude of human-to-human transmission may be underestimated if a longer convalescent phase with higher infectivity is confirmed, or overestimated if infected women cannot transmit Zika virus to their sex partners. We assumed a homogeneous mixing human population, although heterogeneities such as gender, culture, religion, and socioeconomics deserve further investigation. Moreover, it is much more difficult to formulate, parametrize, validate and analyze such models.

If sexual transmission were not small compared to vector-borne transmission, there would have been some excess seen in particular age groups (sexually active age \cite{duffy2009zika}), partnership groups (individuals with regular sexual intercourse), behavioral groups (individuals frequently having unprotected sex), and professional groups (sex workers). A standard questionnaire plus a recent history of sexual contacts for residents in Zika-hit regions will provide these data to examine the mode of sexual transmission.

\vspace*{0.5cm}
\section*{Methods}
\noindent
{\bf Mathematical Model}. \
Human population is divided into six classes: susceptible ($S_h(t)$), exposed ($E_h(t)$), symptomatically infected ($I_{h\!1}(t)$), convalescent ($I_{h\!2}(t)$), asymptomatically infected ($A_h(t)$), and recovered ($R_h(t)$) at time $t>0,$ and the mosquito population is divided into three classes: susceptible ($S_v(t)$), exposed ($E_v(t)$), and infectious ($R_v(t)$), respectively.  $N_h=S_h+E_h+I_{h\!1}+I_{h\!2}+A_h+R_h$ denotes the total number of humans and $N_v=S_v+E_v+I_v$ denotes the total mosquito population, both are assumed to be constant. We use the SEI type of structure for mosquitoes and SEIR type of structure for humans \cite{anderson1991, KeelingRohani}. Based on the assumptions, the ZIKV transmission dynamics between humans and mosquitoes are governed by the following model equations:
\begin{equation*}
%\label{eqn1}
\begin{aligned}
\frac{dS_h}{dt}=& -a b\dfrac{I_v}{N_h} S_h-\beta \dfrac{\kappa E_h+I_{h\!1}+\tau  I_{h\!2}}{N_h}S_h,\\
\frac{dE_h}{dt}=& \theta\Big(a b\dfrac{I_v}{N_h} S_h+\beta \dfrac{\kappa E_h+I_{h\!1}+\tau  I_{h\!2}}{N_h}S_h\Big)-\nu_h E_h,\\
\frac{dI_{h\!1}}{dt}=& \nu_h E_h-\gamma_{h\!1} I_{h\!1},\\
\frac{dI_{h\!2}}{dt}=& \gamma_{h\!1}I_{h\!1}-\gamma_{h\!2} I_{h\!2},\\
\frac{dA_h}{dt}=& (1-\theta)\Big(a b\dfrac{I_v}{N_h} S_h+\beta \dfrac{\kappa E_h+I_{h\!1}+\tau  I_{h\!2}}{N_h}S_h\Big)-\gamma_h A_h,\\
\frac{dR_h}{dt}=& \gamma_{h\!2} I_{h\!2}+\gamma_h A_h,\\
\frac{dS_v}{dt}=& \mu_v N_v -ac \dfrac{ \eta E_h + I_{h\!1}}{N_h}S_v-\mu_v S_v,\\
\frac{dE_v}{dt}=& ac \dfrac{ \eta E_h + I_{h\!1}}{N_h}S_v-(\nu_v+\mu_v) E_v,\\
\frac{dI_v}{dt}=& \nu_v E_v-\mu_v I_v.\\
\end{aligned}
\end{equation*}

\vspace*{0.3cm}
\noindent
{\bf Basic reproduction number}. \
To understand the transmission dynamics, we analyzed the basic reproduction number $\R_0$, which is defined as the number of secondary cases produced by a single infective individual during his/her infectious period in an otherwise wholly susceptible population \cite{anderson1991}. $\R_0$ is a central concept in measuring the transmission of infectious diseases.  For our Zika model, it is given by
\[
\R_0=\frac{\R_{hh}+\sqrt{\R_{hh}^2+4\R_{hv}^2}}{2},
\]
where
\[
\R_{hh}=\frac{\kappa\theta\beta}{\nu_h}+\frac{\theta\beta}{\gamma_{h\!1}}
+\frac{\tau\theta\beta}{\gamma_{h\!2}}\ \mbox{and}\ \R_{hv}=\sqrt{\bigg(\frac{a^2b\eta c m\theta}{\nu_h \mu_v }+\frac{a^2bc m\theta}{ \gamma_{h\!1}\mu_v }\bigg)\frac{\nu_v}{\nu_v+\mu_v}}
\]
are the basic reproduction numbers of sexual transmission and vectorial transmission, respectively.
In particular, $\R_0\approx\R_{hv}\approx\sqrt{\frac{a^2bc m}{\gamma_{h\!1}\mu_v}}$ provides the basic  reproduction number of the classical Ross-Macdonald malaria model \cite{anderson1991} when sexual transmission and asymptomatic infection are rare, and intrinsic and extrinsic incubation periods of ZIKV in humans and mosquitoes are negligible. The necessary and sufficient condition for $\R_0>1$ is $\R_{hh}+\R_{hv}^2>1$, so the percentage of contribution by sexual transmission in $\R_0$ is $\R_{hh}/(\R_{hh}+\R_{hv}^2)$.
%Obviously, $\max\{\R_{hh},\R_{hv}\} \le \R_0 \le \R_{hh}+\R_{hv}$.
$\R_0$ increases when the sexual transmission route is
included. $\R_0$ is increasing in $a, b, c, \eta, m, \beta, \kappa, \tau, \theta, \nu_v$, and decreasing in $\nu_h, \gamma_{h\!1}, \gamma_{h\!2}, \mu_v$.  In particular, $\beta$ (affected by safe-sex education), $m$ (affected by vector control), and $a$ (affected by bed net and insect repellent) are controllable parameters.  Note that $\beta$ is proportional to the fraction of unprotected sexual contact.

\vspace*{0.3cm}
\noindent
{\bf Plug-and-play inference framework}. \
The monthly report cases are obtained as $Z_i=\int_{t_i}^{t_{i+1}}\rho \gamma_{h1} I_{h1}dt,$
where $\rho$ denotes reporting ratio. We assume that the observed monthly cases (both confirmed and suspected) $C_i$ is a random sample from a Negative-binomial (NB) distribution
$C_i \sim \text{NB}\left(n=\frac{1}{\tau_1}, p=\frac{1}{1+Z_i\tau_1}\right),$
where $n$ and $p$ denote the size and probability of the NB distribution, and $\tau_1$ denotes an over-dispersion parameter which will be estimated.
The mean and variance of the NB distribution are given by $\frac{n(1-p)}{p}=Z_i$ and 
$\frac{n(1-p)}{p^2}=Z_i(1+Z_i\tau_1),$ respectively. 
The likelihood function for one country is $L(\theta_1|C_{1,...,N})=\prod\limits_{i=1}^N\,l_i, $
where $\theta_1$ denotes the parameter vector and $l_i$ is the density associated with $C_i$ and $Z_i$.
We apply the framework to three countries -- Brazil, Colombia and El Salvador, which reported most cases continuously, and we denote the overall likelihood for the three countries as
$L_\text{overall}= \prod\limits_{j=1}^3\,L_j.$ 

We use iterated filtering based plug-and-play inference framework \cite{Ioni+06,Ioni+11,He+11,He+10} to estimate $\theta_1$ via maximizing $L_\text{overall}$. Also, we need the following assumptions to keep our model as simple as possible. First, many parameters in the model could be time-dependent. For simplicity, we focus on the scenario that the mosquito-human population ratio $m$ is varying over time and use a cubic spline function to model  $m(t)$. Second, we assume that the three countries share common parameter values, except for country population size and initial conditions. Third, for parameters listed in Table 1, we use parameter values given there except for $m$, which we assume a cubic spline function of time with $n_m$ parameters. Finally, with the two additional parameters, reporting rate $\rho$ and over-dispersion parameter $\tau_1$, the number of parameters ($N_p$) to be fitted is $n_m+2$. We use Bayesian Information Criterion
$BIC= -2 \log L + N_p\log N_d$
to measure goodness-of-fit for models, where $N_d$ denotes number of data points and $N_p$ denotes number of free parameters.

\vspace*{0.3cm}
\noindent
{\bf Acknowledgements} \
Research was partially supported by the Models of Infectious Disease Agent Study (MIDAS) (NIGMS U01GM087728), Early Career Scheme from Hong Kong Research Grants Council (PolyU 251001/14M and PolyU 253004/14P), National Science Foundation (DMS-1518529, DMS-1412454), NSF grant 1414374 as part of the joint NSF-NIH-USDA Ecology and Evolution of Infectious Diseases program, UK Biotechnology
and Biological Sciences Research Council grant BB/M008894/1, Program for Professor of Special Appointment (Eastern Scholar) at Shanghai Institutions of Higher Learning (TP2015050), and Shanghai Gaofeng Project for University Academic Development Program.

\vspace*{0.3cm}
\noindent
{\bf Author Contributors} \
D.G., Y.L. and S.R. developed the model structure; D.G., Y.L. and D.H. performed the modeling and data
analyses; D.H. developed the numerical and statistical analyses; all authors discussed the results and contributed to the writing of the manuscript.

\vspace*{0.3cm}
\noindent
{\bf Declaration of interests} \ 
The authors declare no competing financial interests.


\begin{thebibliography}{99}

\bibitem{duffy2009zika}
Duffy, M. R., Chen, T. H., Hancock, W. T.,  \textit{et al.} Zika virus outbreak on Yap Island, Federated States of Micronesia, \textit{New Engl. J. Med.} \textbf{360}(24), 2536--2543, (2009).

\bibitem{Mlakar2016}
Mlakar, J., Korva, M., Tul, N., \textit{et al.} Zika virus associated with microcephaly, \textit{New Engl. J. Med.} Published online February 10, 2016. DOI: 10.1056/NEJMoa1600651.

\bibitem{Cauchemez2016}
Cauchemez, S., Besnard, M., Bompard, P., \textit{et al.} Association between Zika virus and microcephaly in French Polynesia, 2013–15: a retrospective study,  \textit{Lancet} Published online March 15, 2016.
http://dx.doi.org/10.1016/S0140-6736(16)00651-6.

\bibitem{Cao-Lormeau2016}
Cao-Lormeau, V.-M., Blake, A., Mons, S., \textit{et al.} Guillain-Barr\'e Syndrome outbreak associated with Zika virus infection in French Polynesia: a case-control study, \textit{Lancet} Published online February 29, 2016. http://dx.doi.org/10.1016/S0140-6736(16)00562-6.

\bibitem{dick1952zika}
Dick, G. W., Kitchen, S. F., Haddow, A. J. Zika virus. I. Isolations and serological specificity, \textit{Trans. R. Soc. Trop. Med. Hyg.} \textbf{46}(5), 509--520, (1952).

\bibitem{macnamara1954zika}
Macnamara, F. N. Zika virus: a report on three cases of human infection during an epidemic of jaundice in {Nigeria}, \textit{Trans. R. Soc. Trop. Med. Hyg.} \textbf{48}(2), 139--145, (1954).

\bibitem{Shapshak2015}
Shapshak, P., Somboonwit, C., Foley, B. T.,  \textit{et al.} Zika virus, in \textit{``Global Virology I - Identifying and Investigating Viral Diseases''} (eds.  P. Shapshak et al.)  Springer, 477-500 (New York, 2015).

\bibitem{musso2014rapid}
Musso, D., Nilles, E. J., Cao-Lormeau, V.-M. Rapid spread of emerging Zika virus in the Pacific area, \textit{Clin. Microbiol. Infect.} \textbf{20}(10), 595--596, (2014).

\bibitem{PAHO2016}
Pan American Health Organization (PAHO), Zika virus infection. 
http://www.paho.org/hq/index.php$?$option=com$_-$content\&view=article\&id=11585\newline
\&Itemid=41688\&lang=en (Accessed on February 26, 2016).

\bibitem{WHO2016}
World Health Organization (WHO), WHO statement on the first meeting of the International Health Regulations (2005) Emergency Committee on Zika virus and observed increase in neurological disorders and neonatal malformations, February 1, 2016.
http://www.who.int/mediacentre/news/statements/2016/1st-emergency-committee-zika/en/
 (Accessed on February 26, 2016).

\bibitem{CDC2016}
Centers for Disease Control and Prevention (CDC), CDC Emergency Operations Center moves to highest level of activation for Zika response, Febraury 3, 2016. 
http://www.cdc.gov/media/releases/2016/s0208-zika-eoca-activation.html
(Accessed on February 26, 2016).

\bibitem{gourinat2015detection}
Gourinat, A. C., O'Connor, O., Calvez, E., Goarant, C., Dupont-Rouzeyrol, M.
Detection of {Zika} virus in urine, \textit{Emerg. Infect. Dis.} \textbf{21}(1), 84-86, (2015).

\bibitem{musso2015detection}
Musso, D., Roche, C., Nhan, T. X., \textit{et al.} Detection of {Zika} virus in saliva, \textit{J. Clin. Virol.} \textbf{68}, 53--55, (2015).

\bibitem{Atkinson2016}
Atkinson, B., Hearn, P., Afrough, B., \textit{et al.} Detection of Zika virus in semen, \textit{Emerg. Infect. Dis.} http://dx.doi.org/10.3201/eid2205.160107, (2016).

\bibitem{musso2015potential}
Musso, D., Roche, C., Nhan, T. X., \textit{et al.} Potential sexual transmission of {Zika} virus, \textit{Emerg. Infect. Dis.} \textbf{21}(2), 359--361, (2015).

\bibitem{foy2011probable}
Foy, B. D., Kobylinski, K.C., Foy, J. L., \textit{et al.} Probable non--vector-borne transmission of Zika virus, Colorado, USA, \textit{Emerg. Infect. Dis.} \textbf{17}(5), 880--882, (2011).

\bibitem{dchhs2016first}
Dallas~County Health and Human Services (DCHHS),
\newblock {DCHHS Reports First Zika Virus Case in Dallas County Acquired Through Sexual Transmission}, February 2, 2016.

\bibitem{Hills2016}
Hills, S. L., Russell, K., Hennessey, M., \textit{et al.} Transmission of Zika virus through sexual contact with travelers to areas of ongoing transmission -- Continental United States, 2016. \textit{MMWR Morb. Mortal. Wkly. Rep.} ePub: 26 February, 2016. DOI: http://dx.doi.org/10.15585/mmwr.mm6508e2er.

\bibitem{Mansuy2016}
Mansuy, J.M., Dutertre, M., Mengelle, C., \textit{et al.} Zika virus: high infectious viral load in semen, a new sexually transmitted pathogen? \textit{Lancet Inf. Dis.} Published online March 3, 2016. http://dx.doi.org/10.1016/S1473-3099(16)00138-9

%\bibitem{DailyMail}
%Mezzofiore, G. Europe's first case of sexually transmitted Zika virus is diagnosed in France as a woman %falls ill after her partner travelled to South America, \textit{Daily Mail}, February 27, 2016. %http://www.dailymail.co.uk/news/article-3467037/First-case-sexually-transmitted-Zika-virus-diagnosed-%France-woman-falls-ill-partner-travelled-South-America.html

\bibitem{Venturi2016}
Venturi, G., Zammarchi, L., Fortuna, C. \textit{et al} An autochthonous case of Zika due to possible sexual
transmission, Florence, Italy, 2014, \textit{Euro Surveill.} 2016; \textbf{21}(8), pii:30148. 
DOI: http://dx.doi.org/10.2807/1560-7917.

%\bibitem{oehler2014zika}
%Oehler, E. \textit{et al.} Zika virus infection complicated by Guillain-Barr{\'e} syndrome-case report, %French Polynesia, December 2013, \textit{Euro. Surveill.} \textbf{19}(9), pii: 20720 (2014).

%\bibitem{schuler2016possible}
%L. Schuler-Faccini, E. M. Ribeiro, I. M. Feitosa, D. D. Horovitz, D. P. Cavalcanti, A. Pessoa, M. J. Doriqui, %J. I. Neri, J. M. Neto, H. Y. Wanderley, M. Cernach, A. S. El-Husny, M. V. Pone, C. L. Serao, M. T. %Sanseverino, Brazilian Medical Genetics Society-Zika Embryopathy Task Force, Possible association %between Zika virus infection and microcephaly -- Brazil, 2015, \textit{Morb. Mortal. Wkly. Rep.}, %\textbf{65}(3), 59-62 (2016).

%\bibitem{cao2014zika}
%V. M. Cao-Lormeau, C. Roche, A. Teissier, E. Robin, A. L. Berry, H. P. Mallet, A. A. Sall, D. Musso, Zika %virus, French Polynesia, South Pacific, 2013, \textit{Emerg. Infect. Dis.}, \textbf{20}(6), 1085-1086 (2014).

%\bibitem{musso2014potential}
%Musso, D. \textit{et al.}  Potential for Zika virus transmission through blood transfusion demonstrated %during an outbreak in French Polynesia, November 2013 to February 2014, \textit{Euro. Surveill.} %\textbf{19}(14), pii: 20761 (2014).

%\bibitem{besnard2014evidence}
%Besnard, M., Last{\`e}re, S., Teissier, A., Cao-Lormeau, V. M. \& Musso, D. Evidence of perinatal %transmission of Zika virus, French Polynesia,  December 2013 and February 2014, \textit{Euro Surveill.} %\textbf{19}(13), pii: 20751 (2014).

%\bibitem{fonseca2014first}
%K. Fonseca, B. Meatherall, D. Zarra, M. Drebot, J. MacDonald, K. Pabbaraju, S. Wong, P. Webster, R. %Lindsay, R. Tellier, First case of {Zika} virus infection in a returning {Canadian} traveler, \textit{Am. J. Trop. %Med.}, \textbf{91}(5), 1035--1038 (2014).

%\bibitem{kutsuna2014two}
%S. Kutsuna, Y. Kato, T. Takasaki, et al., Two cases of Zika fever imported from French Polynesia to %Japan, December 2013 to January 2014, \textit{Euro Surveill}, \textbf{19}(4), pii: 20683 (2014).

\bibitem{CDC2016b}
Centers for Disease Control and Prevention (CDC), Update: Interim Guidelines for Prevention of Sexual Transmission of Zika Virus -- United States, 2016, Febraury 23, 2016. http://emergency.cdc.gov/han/han00388.asp

%\bibitem{Oster2016}
%A. M. Oster, J. T. Brooks, J. E. Stryker, et al. Interim guidelines for prevention
%of sexual transmission of Zika virus -- United States, 2016, \textit{MMWR Morb. Mortal. Wkly. Rep.} %\textbf{65} (Early Release): 1-2 (2016). DOI: http://dx.doi.org/10.15585/mmwr.mm6505e1e.

%\bibitem{duong2015asymptomatic}
%Duong, V. \textit{et al.} Asymptomatic humans transmit dengue virus to mosquitoes, \textit{Proc. Natl. %Acad. Sci. USA} \textbf{112}(47), 14688--14693 (2015).

\bibitem{anderson1991}
Anderson, R. M., May, R. M. \textit{Infectious Diseases of Humans: Dynamics and Control}. Oxford University Press (Oxford, 1991).

\bibitem{KeelingRohani} 
Keeling, M. J., Rohani, P. {\it Modeling Infectious Diseases in Humans and Animals}, Princeton University Press (Princeton, 2007).

\bibitem{craig-gray-edwards2015}
Craig, A. P., Gray, R. T., Edwards, J. L., \textit{et al.} The potential impact of vaccination of the prevalence of gonorrhea, \textit{Vaccine} \textbf{33}(36), 4520--4525, (2015).

\bibitem{wald2001effect}
Wald, A.,  Langenberg, A. G., Link, K., \textit{et al.} Effect of condoms on reducing the transmission of herpes simplex virus type 2 from men to women, \textit{JAMA} \textbf{285}(24), 3100--3106, (2001).

\bibitem{laumann1994social}
Laumann, E. O., Gagnon, J. H.,  Michael, R. T. \& Michaels, S. \textit{The Social Organization of Sexuality: Sexual Practices in the United States}, University of Chicago Press (Chicago, 1994).

\bibitem{andraud2012dynamic}
Andraud, M., Hens, N., Marais, C., Beutels, P. Dynamic epidemiological models for dengue transmission: a systematic review of structural approaches, \textit{PLoS One} \textbf{7}(11), e49085, (2012).

\bibitem{chikaki2009dengue}
Chikaki, E., Ishikawa, H. A dengue transmission model in {Thailand} considering sequential infections with all four serotypes, \textit{J. Infect. Dev. Countr.} \textbf{3}(9), 711--722, (2009).

\bibitem{de2011modeling}
de Castro Medeiros, L. C., Castilho, C. A., Braga, C., \textit{et al.} Modeling the dynamic transmission of dengue fever: investigating disease persistence, \textit{PLoS. Negl. Trop. Dis.} \textbf{5}(1), e942, (2011).

\bibitem{bearcroft1956zika}
Bearcroft, W. G. C. Zika virus infection experimentally induced in a human volunteer,
\textit{Trans. R. Soc. Trop. Med. Hyg.} \textbf{50}(5), 442--448, (1956).

\bibitem{boorman1956simple}
Boorman, J. P., Porterfield, J. S. A simple technique for infection of mosquitoes with viruses transmission of {Zika} virus, \textit{Trans. R. Soc. Trop. Med. Hyg.} \textbf{50}(3), 238--242, (1956).

\bibitem{Ioni+06}
Ionides, E. L., Bret{\'o}, C., King, A. \newblock Inference for nonlinear dynamical systems.
\newblock {\it Proc. Natl. Acad. Sci. USA}  {\bf 103},18438-18443, (2006).

\bibitem{Ioni+11}
Ionides, E. L., Bhadra, A., Atchad\'e, Y., \textit{et al.} \newblock Iterated filtering. \newblock {\it Ann. Stat.} {\bf 39}, 1776--1802, (2011).

%\bibitem{Earn+12}
%Earn, D. J. D., He, D. H., Loeb, M. B., Fonseca, K., Lee, B. E., Dushoff, J. \newblock Effects of school %closure on incidence of pandemic influenza in  Alberta, Canada. \newblock {\it Ann. Intern. Med.} 2012; %{\bf 156}, 173--181.

%\bibitem{Cama+11}
%Camacho, A., Ballesteros, S., Graham, A. L., Carrat, F., Ratmann, O., Cazelles, B.
%\newblock Explaining rapid reinfections in multiple-wave influenza outbreaks: Tristan da Cunha 1971 %epidemic as a case study. \newblock {\it Proc. Roy. Soc. B}  2011; {\bf 278}, 3635--3643.

\bibitem{He+11}
He, D. H., Dushoff, J., Day, T., Ma, J. L., Earn, D. J. D.
\newblock Mechanistic modelling of the three waves of the 1918 influenza pandemic.
\newblock {\it Theor. Ecol}.  {\bf 4}, 283--288, (2011).

\bibitem{He+10}
He, D. H., Ionides, E. L.,  King, A. A.
\newblock Plug-and-play inference for disease dynamics: measles in large and small populations as a case study. \newblock {\it J. R. Soc. Interface} {\bf 7}, 271--283, (2010).

%\bibitem{King+08}
%King, A. A., Ionides, E. L., Pascual, M., Bouma, M. J.
%\newblock Inapparent infections and cholera dynamics.
%\newblock {\it Nature} 2008; {\bf 454}, 877--880.

\bibitem{iman-helton-campbell1981}
Iman, R. L., Helton, J. C., Campbell, J. E. An approach to sensitivity analysis of computer models: I - introduction, input variable selection and preliminary variable assessment, \textit{J. Qual. Technol.} \textbf{13}(3), 174--183, (1981).

\bibitem{saltelli2000sensitivity}
Saltelli, A., Chan, K., Scott, E. M. \textit{Sensitivity Analysis}, Wiley (Chichester, 2000).

\bibitem{ma2006generality}
Ma J., Earn, D. J. Generality of the final size formula for an epidemic of a newly invading infectious disease, \textit{Bull. Math. Biol.} \textbf{68}(3), 679--702, (2006).

\bibitem{marques1994basic}
Marques, C. A., Forattini, O. P., Massad, E. The basic reproduction number for dengue fever in {S\~{a}o Paulo state, Brazil}: 1990--1991 epidemic, \textit{Trans. R. Soc. Trop. Med. Hyg.} \textbf{88}(1), 58--59, (1994).

%\bibitem{Heesterbeek2015}
%H. Heesterbeek, R. M. Anderson, V. Andreasen, et al.,  Modeling infectious disease dynamics in the %complex landscape of global health, \textit{Science}, \textbf{347}(6227), aaa4339 (2015).

\bibitem{Colgate1989}
Colgate, S. A., Stanley, E. A., Hyman, J. M., Layne, S. P., Qualls, C.  Risk behavior-based model of the cubic growth of acquired immunodeficiency syndrome in the United States,  \textit{Proc. Natl. Acad. Sci. USA} \textbf{86} (12), 4793-4797, (1989).

\end{thebibliography}
\end{document}